\documentclass[11pt]{IEEEtran}
\usepackage{graphicx}
\usepackage[hidelinks]{hyperref}
\usepackage[backend=biber,style=authoryear,sorting=nyt,url=false, maxcitenames=2,natbib,uniquename=false,uniquelist=false,giveninits=true]{biblatex}
\usepackage{amsmath}
\bibliography{bibliography}

\AtEveryBibitem{\clearfield{month}}
\AtEveryBibitem{\clearfield{day}}
\AtEveryBibitem{\clearfield{issn}}

\DeclareCiteCommand{\citeyear}
    {}
    {\bibhyperref{\printdate}}
    {\multicitedelim}
    {}

\begin{document}

\title{Modelling the influence of streamwise flow field acceleration on the aerodynamic performance of an actuator disc}

\author{\IEEEauthorblockN{Clemens Paul Zengler\IEEEauthorrefmark{1},
Niels Troldborg, Mac Gaunaa.}\\
\IEEEauthorblockA{Department of Wind and Energy Systems,
Technical University of Denmark.\\
\mbox{Frederiksborgvej 399, 4000 Roskilde},
Denmark.\\
Email: \IEEEauthorrefmark{1}clezen@dtu.dk}}

\maketitle

\begin{abstract}
Streamwise acceleration of the background flow field is one of various effects occurring when wind turbines operate under non-idealized conditions, such as in complex terrain or in dense wind farms. Thus, studying this effect is essential to improve understanding of aerodynamic performance in these cases. 
In the present work, a simple model based on momentum theory is derived for the situation of an actuator disc (AD) operating in a background flow field with a constant velocity gradient. Reynolds-averaged Navier-Stokes (RANS) simulations of this scenario are performed, showing that a positive acceleration yields a reduction of induction and vice versa, a negative acceleration leads to an increase of induction. The new model accurately captures this behavior and reduces the prediction error by eighty percent compared to classical momentum theory where the effect of the background flow acceleration is disregarded. Further analysis suggests that the model can be extended to a formulation in which an even higher prediction accuracy can be achieved. 
\end{abstract}

\section{Introduction}
\noindent Classical momentum theory as developed by \textcite{rankine_MechanicalPrinciplesAction_1865}, \textcite{froude_ElementaryRelationPitch_1878}, and \textcite{froude_PartPlayedPropulsion_1889} yields insights into the relation between thrust and induction of an idealized wind turbine operating in a uniform flow field. Also, the theoretical limit of aerodynamic performance, usually referred to as the Betz limit, can be derived from it \citep{betz_MaximumTheoretischMoglichen_1920}. The aerodynamic performance referred to here is the aerodynamic power, which is a product of thrust and velocity in the turbine plane.
By today, momentum theory is a fundamental part of the aeroelastic design of wind turbines and power performance predictions. 
However, due to its simplifying assumptions which do not necessarily apply to the actual operating conditions of modern wind turbines, various modifications were proposed in the past. 
When turbines operate in the atmospheric boundary layer, they are usually subject to wind shear, thus a variation of streamwise velocity over height. To account for this, several models were proposed in the past. \textcite{wagner_AccountingSpeedShear_2011} suggest using a reference velocity for performance predictions based on the kinetic energy flux through the turbine disc plane and validated this model with measurements of a full-scale turbine. Later, \textcite{chamorro_NonuniformVelocityDistribution_2013} derived an analytical correction of the Betz limit by including the effect of shear upstream and downstream of the wind turbine in the model. Since shear would also lead to a variation of thrust force along the disc, \textcite{draper_PerformanceIdealTurbine_2016} included this effect in their model. 
Based on vortex theory, \textcite{gaunaa_SimpleVortexModel_2023} also showed that momentum theory should be applied locally within momentum-based design and analysis tools to properly account for the effect of shear. 

The problem of yaw misalignment and its effect on induction was approached in the early days by \textcite{glauert_GeneralTheoryAutogyro_1926} and more recently by  \textcite{heck_ModellingInductionThrust_2023} and then by \textcite{tamaro_PowerControlMisaligned_2024}, who incorporate both, shear and yaw misalignment in their model. 

\textcite{mikkelsen_ModellingWindTunnel_2002} developed a correction to account for the effect of wind tunnel blockage on the aerodynamic performance of turbines in wind tunnels.  The potential for using a diffuser to improve the aerodynamic performance of turbines was for example studied by \textcite{jamieson_BeatingBetzEnergy_2009}. \textcite{sorensen_GeneralMomentumTheory_2016a} discusses the theory of both, wind tunnel blockage and diffuser modelling, in more detail.

In complex terrain or in dense wind farms, wind turbines may be subject to streamwise variations of the background flow field. Various authors have shown that these variations can have a significant impact on the aerodynamic performance of a wind turbine. 

\textcite{troldborg_BriefCommunicationHow_2022} investigated the power performance of a full-scale wind turbine on the ridge of a hill through Large Eddy Simulations (LES) and found that depending on the streamwise development of the undisturbed flow field behind the ridge, significant variations in performance could be measured. Similar results were found in an experimental campaign by \textcite{dar_ExperimentalAnalyticalStudy_2023}. Most recently, \textcite{zengler_FreeWindSpeed_2024}, and later in the same year \textcite{revaz_EffectHillsWind_2024}, systematically performed simulations of wind turbines on idealized hills which confirmed the effect of streamwise flow variations on aerodynamic performance.
Further works in which the influence of streamwise flow variations on power performance is reported exist (see e.g. \textcite{cai_LocalTopographyinducedPressure_2021} and \textcite{mishra_WakeSteeringWind_2024}). However, it is often the case that the effect is not investigated in isolation, making it difficult to draw conclusions. 
In general, most research shows that negative flow acceleration (deceleration) behind the turbine can lead to a decrease of performance, while positive flow acceleration results in an increase of performance.

Efforts to include the effect of acceleration in turbine modelling have been mainly focused on the wake behavior under such circumstances \citep{shamsoddin_ModelEffectPressure_2018, dar_AnalyticalModelWind_2022, dar_ExperimentalAnalyticalStudy_2023}. 
Only \textcite{cai_LocalTopographyinducedPressure_2021} attempted to also model power performance in the presence of pressure gradients. For this purpose, they used a wake model developed for pressure gradients in combination with a linearized flow solver and a control volume analysis.
Although this model can predict both wake profiles and power performance which are in sound agreement with measurements, their approach comes with a computational burden and also cannot be directly incorporated into aeroelastic design as it does not yield local information about the flow state in the turbine plane.

Despite these efforts, no further work is known to the authors, in which an analytical model was developed that accounts for the effect of streamwise flow acceleration on the aerodynamic performance. The present work is an attempt to model the effect of a constant streamwise flow acceleration on the induction and thus also on the aerodynamic performance of a wind turbine. 

The remainder of this article is structured as follows.
First, the simplified model to account for the effect of a streamwise flow acceleration in momentum theory is derived. 
Second, the model is validated with Reynolds-averaged Navier-Stokes (RANS) simulations of an actuator disc (AD) exposed to various constant velocity gradient flows. 
Last, implications and modelling details are discussed.  

\section{Model derivation}
\begin{figure}[t]
    \centering
    \includegraphics[width=8.5cm]{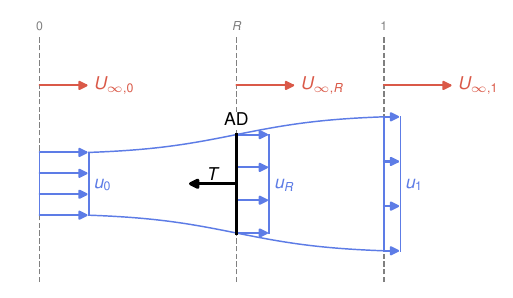}
    \caption{Sketch of the problem and notation for theoretical analysis.}
    \label{fig:fig_1}
\end{figure}
\noindent A steady, incompressible, inviscid, and one-dimensional flow past an AD is assumed.
Figure \ref{fig:fig_1} shows a sketch of the problem and its notation, which is based on the following logic. $U_\infty$ is the undisturbed constant velocity gradient flow field, with $_\infty$ referring generally to quantities in the undisturbed flow. $u$ is the disturbed flow field. The subscript $_0$ denotes quantities ahead of the disc plane, $_R$ denotes quantities in the disc plane, and $_1$ denotes the ultimate wake. Further, the subscript $_u$ denotes the case of a uniform flow field without a velocity gradient. 

The undisturbed flow field is subject to a constant velocity gradient, thus 
\begin{align}
    \frac{\operatorname{d} U_\infty}{\operatorname{d} x} = C .
\end{align}
By the virtue of Bernoulli's equation, the pressure jump $\Delta p$ across the AD fulfills 
\begin{align}
\Delta p  = \frac{1}{2}\rho U_{\infty,0}^2 + p_{\infty,0} - \frac{1}{2}\rho u_1^2-p_1. 
\end{align}
Further, the undisturbed flow quantities ahead and behind the disc are related by Bernoulli's theorem as 
\begin{align}
\frac{1}{2}\rho U_{\infty,0}^2 + p_{\infty,0} = \frac{1}{2}\rho U_{\infty,1}^2 + p_{\infty,1},
\end{align}
which yields for the pressure jump
\begin{align}
\Delta p  = \frac{1}{2}\rho U_{\infty,1}^2 + p_{\infty,1} - \frac{1}{2}\rho u_1^2-p_1. 
\end{align}
By assuming, as in classical momentum theory 
(\citeauthor{rankine_MechanicalPrinciplesAction_1865}, \citeyear{rankine_MechanicalPrinciplesAction_1865}; \citeauthor{froude_ElementaryRelationPitch_1878}, \citeyear{froude_ElementaryRelationPitch_1878}; \citeauthor{froude_PartPlayedPropulsion_1889}, \citeyear{froude_PartPlayedPropulsion_1889}),
that the pressure difference between the wake and the free stream is zero, the pressure drop over the AD is seen to represent the ultimate difference in kinetic energy between free stream and wake
\begin{align}\label{eq:dp1}
\Delta p  = \frac{1}{2}\rho \left(U_{\infty,1}^2 -u_1^2\right). 
\end{align}
Until now, no assumptions about the onset of the ultimate wake were made. For the undisturbed flow, the onset is related to the velocity in the AD plane by 
\begin{align}\label{eq:u1inf}
    U_{\infty,1} = U_{\infty,R} + L \frac{\operatorname{d} U_\infty}{\operatorname{d} x}, 
\end{align}
with the unknown length scale $L$. One of the key elements in the present model is the assumption that the wake flow is affected by the acceleration to the same extent as the free stream
\begin{align}\label{eq:u1}
    u_1 = u_{u,1} + L \frac{\operatorname{d} U_\infty}{\operatorname{d} x}, 
\end{align}
where $u_{u,1}$ denotes the wake velocity in a uniform flow field. Combining Eq. \eqref{eq:u1} and Eq. \eqref{eq:u1inf} with \mbox{Eq. \eqref{eq:dp1}} and rearranging results in 
\begin{align}\label{eq:dp2}
\Delta p  = \frac{1}{2}\rho \left(U_{\infty,R}^2 -u_{u,1}^2\right) + \rho L \left(U_{\infty,R} - u_{u,1}\right)\frac{\operatorname{d} U_\infty}{\operatorname{d} x}.
\end{align}
Lastly, the fact that for a uniform flow
\begin{align}
2a = 1-\frac{u_{u,1}}{U_{\infty,R}}
\end{align}
is valid with the induction factor generally defined as 
\begin{align}
    a = 1- \frac{u_{R}}{U_{\infty,R}}
\end{align}
is applied. Also, the pressure jump is replaced by the thrust divided by the AD area $\Delta p = \frac{T}{A}$ and all quantities are normalized by $\frac{1}{2}\rho U_{\infty,R}^2$ which yields the final relation between induction and thrust in an accelerating background flow field
\begin{align}\label{eq:fe}
    C_T = 4 a (1-a) + 4 a l \beta
\end{align}
with the thrust coefficient $C_T$, the non-dimensional length scale $l = L/D$ and the non-dimensional velocity gradient $\beta = \frac{D}{U_{\infty,R}}\frac{\operatorname{d} U_\infty}{\operatorname{d} x}$ where D denotes the AD diameter. From this equation, it can be seen that the relation between induction and thrust in a uniform flow field is modified by an additive term proportional to the free stream velocity gradient in case of an accelerating flow field. The length scale $l$ is an unknown quantity.   

Since the first component of the right-hand side of Eq.~\eqref{eq:fe} simply resembles the thrust-induction relation in a uniform flow, a reasonable generalization might be that arbitrary thrust-induction curves obtained in a uniform flow field can be corrected for acceleration by simply adding the last therm of the right-hand side of Eq. \eqref{eq:fe}. Thus
\begin{align}\label{eq:totct}
    C_T(a) = C_{T,u}(a) + 4 a l \beta,
\end{align}
with the thrust-induction relation $C_{T,u}(a)$ for a uniform background flow field. Equation \eqref{eq:totct} is the new model, which will be validated in the next section. 

\section{Model validation}
\noindent The derived model is validated by performing three-dimensional RANS simulations of an AD operating in a flow field featuring a constant velocity gradient in the streamwise direction. The thrust-induction curves for a wide range of positive and negative accelerations are evaluated. 
\begin{figure}[t]
    \centering
    \includegraphics[width=8.5cm]{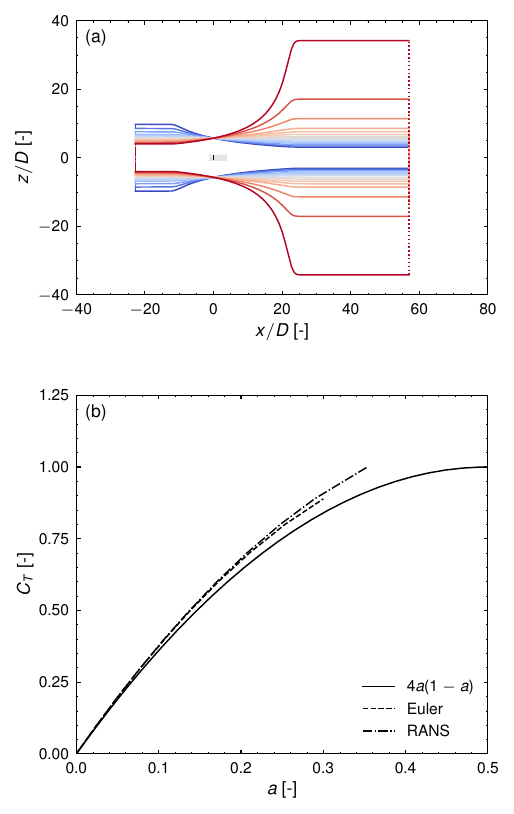}
    \caption{\textbf{(a)} Boundaries of the computational domains considered simulating a constant velocity gradient. In the lateral direction, the domain width is 11.390~D. Dash-dotted lines indicate the inlet, dotted lines the outlet, and solid lines the vertical boundaries. The gray area indicates a region of mostly cubic cells of a side length of approximately 0.047~D. The color-coding for the different domains is used consistently in all subsequent figures.
    \textbf{(b)} Comparison of Eulerian and RANS simulation for the baseline uniform case with theoretical results from momentum theory.}
    \label{fig:fig_2}
\end{figure}
\begin{figure*}[t]
    \centering
    \includegraphics[width=17.7cm]{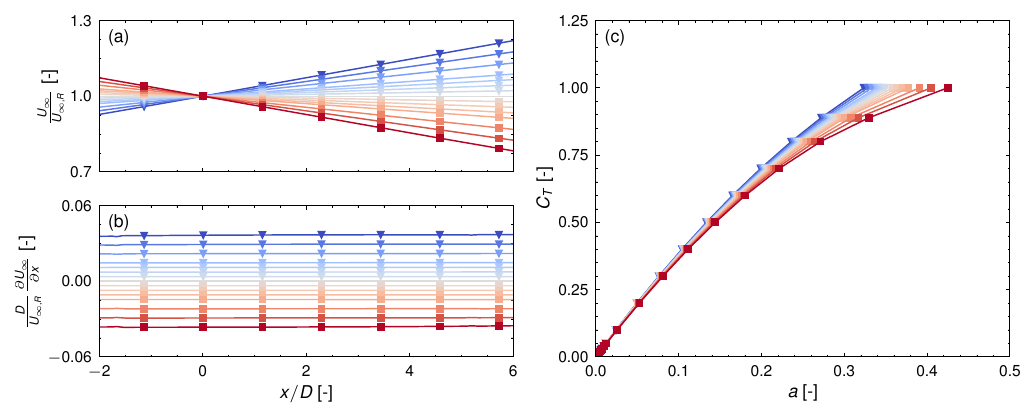}
    \caption{Simulation results with \textbf{(a)} the undisturbed streamwise velocity along the centerline of the AD, \textbf{(b)} the streamwise gradient, and \textbf{(c)} the measured thrust-induction curves.}
    \label{fig:fig_3}
\end{figure*}
\subsection{Grid design}
\noindent For every considered velocity gradient, a separate computational grid is generated.
In all grids, the Cartesian coordinate system is centered around the AD with $x$ denoting the streamwise, $y$ the lateral, and $z$ the vertical direction.
The cross-sectional area $A_S(x)$ of each grid is defined from mass conservation by the constant velocity gradient condition, thus 
\begin{align}\label{eq:geom}
    A_S(x) = \frac{U_R A_{S,R}}{U_R + x \frac{\operatorname{d} U_\infty}{\operatorname{d} x}}, 
\end{align}
with $U_R$ and $A_{S,R}$ being the velocity and the cross-sectional area in the AD plane. These quantities are held constant for all grids. The latter is quadratic with a side length of 11.390~D resulting in a blockage ratio of 0.605~\%\footnote{The seemingly arbitrary value of 11.390~D is the consequence of the radial smearing of the AD thrust force as described by \textcite{zengler_FreeWindSpeed_2024}. If just the radial extent of the 
force distribution was used as reference length, the side length would be 10~D. However, to make the results comparable, the normalization diameter is calculated based on an equivalent AD with a uniform loading over the entire extent of the AD.}. Based on Eq. \eqref{eq:geom}, the upper and lower boundaries of the grids are modified vertically for the different velocity gradients. The lateral boundaries are not modified. Further, inlet and outlet regions are appended to the geometry defined by Eq. \eqref{eq:geom} and the transition between these regions smoothened by a Laplacian smoothing algorithm. 
The vertical boundaries of the domains are visualized in \mbox{Fig. \ref{fig:fig_2} (a)}.
The grids are curvilinear and consist of 320$\times$128$\times$128 cells in $x$, $y$, and $z$ direction, respectively. In the region of the AD, the cells are nearly cubic, with a resolution of 21 cells per diameter. 

\subsection{Actuator disc model}
\noindent The AD is simulated with a uniform force distribution. As described by \textcite{zengler_FreeWindSpeed_2024}, the outer 25~\% of the radial boundaries are linearly smeared out to improve solution convergence. The induction is evaluated as the mean induction over the area of the disc which is not affected by the smearing. 
The actuator shape method is applied to project the force into the computational grid \citep{rethore_VerificationValidationActuator_2014,troldborg_ConsistentMethodFinite_2015}. In Fig. \ref{fig:fig_2} (b), the thrust-induction curves obtained using RANS and Euler equations for a uniform background flow are compared with momentum theory. It is seen, that the Eulerian simulation does not perfectly agree with momentum theory. This discrepancy can be attributed to the numerical discretization as it has been reported in the past \citep{hodgson_QuantitativeComparisonAeroelastic_2021,mikkelsen_ActuatorDiscMethods_2004}. If it is desired to better agree with the momentum theory results, it is necessary to increase the grid resolution or to probe the induction slightly behind the position of the AD. However, within this work, no correction is performed, thus the curve representing the RANS results is used as the zero-acceleration/uniform curve in the following. 

\begin{figure*}[t]
    \centering
    \includegraphics[width=17.7cm]{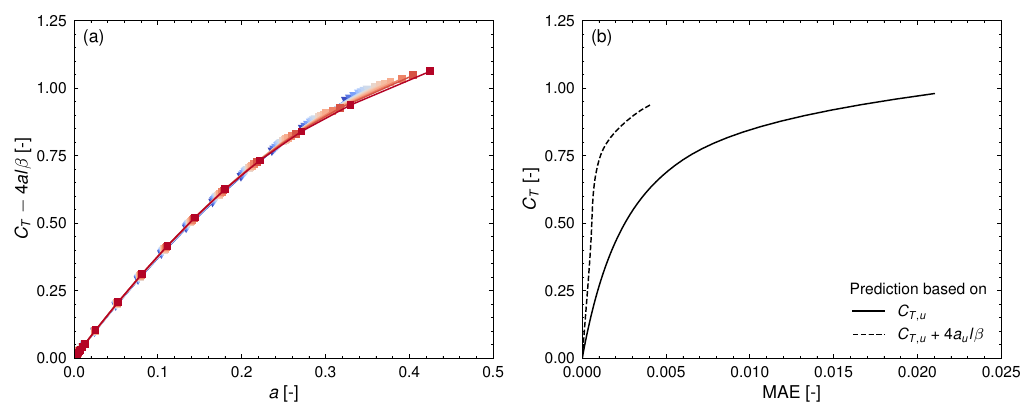}
    \caption{Model predictions for $l = \text{1}$: \textbf{(a)} collapsed induction curves for $C_{T} - 4 a l \beta$ and \textbf{(b)} MAE for induction predictions without and with correction for acceleration.}
    \label{fig:fig_4}
\end{figure*}

\subsection{Boundary conditions, RANS closure and solver}
\noindent Inlet and outlet conditions are applied at the streamwise boundaries. 
The inlet velocity is specified such that $U_R$ is reached in the AD plane. 
The chosen $U_R$ corresponds to a diameter-based Reynolds number of \mbox{3.798 million}. 
Inlet turbulence properties correspond to a turbulence intensity of \mbox{1 \%} with the mixing length specified to be 0.228~D. The lateral boundaries are periodic and slip conditions are applied at the upper and lower boundaries.
As RANS closure, the two-equation $k$-$\omega$-$SST$ model without modification of the standard model coefficients is applied \citep{menter_TwoequationEddyviscosityTurbulence_1994}. The incompressible, three-dimensional, Navier-Stokes solver \mbox{EllipSys3D} \citep{sorensen_GeneralPurposeFlow_1995,michelsen_Basis3DPlatformDevelopment_1992} is employed. The used solution algorithm is a SIMPLE-like procedure based on the non-relaxed momentum equations \citep{sorensen_EllipSys2D3DUser_2018}. Rhie-Chow interpolation is included to avoid decoupling of the pressure and velocity fields \citep{rhie_NumericalStudyTurbulent_1983} and the convective terms are discretized with the QUICK scheme \citep{leonard_StableAccurateConvective_1979}.

\subsection{Simulation results}
\noindent Simulation results are presented in Fig. \ref{fig:fig_3}. The color-coding is used consistently with Fig. \ref{fig:fig_2} (a) and the subsequent Fig. \ref{fig:fig_4} and Fig. \ref{fig:fig_6} where each color represents one specific simulation. 
In panels (a) and (b) of Fig.~\ref{fig:fig_3} the normalized streamwise velocities $U_\infty$ along the centerline and the respective normalized gradients for the different domains presented previously are shown. In all cases, a nearly constant streamwise gradient is obtained ahead and behind the AD. 
Panel (c) shows the measured thrust-induction curves obtained by applying a certain thrust coefficient and evaluating the corresponding induction. 
With an increasing thrust coefficient, the differences between the individual curves grow. In cases of a positive acceleration, the induction is reduced, while it is increased in cases of a negative acceleration. It is interesting to notice, that for a given magnitude of acceleration, a negative acceleration leads to a higher variation in induction than a positive acceleration, which is in agreement with experimental observations from \mbox{\textcite{dar_ExperimentalAnalyticalStudy_2023}}, who saw a higher impact on the power coefficient in case of a negative acceleration than in case of a positive acceleration.

\subsection{Modelling results}
\noindent Based on the thrust-induction curves presented in Fig.~\ref{fig:fig_3}, it is investigated if the model represented by Eq. \eqref{eq:totct} is capable of predicting the observed trends. The non-dimensional velocity gradient $\beta$ is evaluated in the undisturbed flow field at the position of the AD in the center, thus in Fig. \ref{fig:fig_3} (b) at $x/D = \text{0}$. 
Further, the length scale $l$ needs to be determined. 
\textcite{crespo_SurveyModellingMethods_1999} stated that the length of the initial wake expansion region, thus the region where wake and free stream pressure equalize, is one diameter long, a value which was also adopted by \textcite{dar_AnalyticalModelWind_2022}.
Thus, without further justification, $l$ is set to one diameter. The validity of this assumption will be discussed later in Sec. \ref{subsec:theoptimallengthscale}.

Model predictions are presented in Fig. \ref{fig:fig_4}. 
If the model works as expected, solving Eq. \eqref{eq:totct} for $C_{T,u}(a) = C_T(a) -4 a l \beta $ and inserting the respective simulation results into the right-hand side of this formula should lead to the same curve for all simulations. This is visualized in panel (a) of \mbox{Fig. \ref{fig:fig_4}}. Indeed, it can be seen that all curves previously shown in \mbox{Fig. \ref{fig:fig_3} (c)} collapse consistently on the zero-acceleration curve.

To quantify the model accuracy, the mean absolute error (MAE) is evaluated as
\begin{equation}
    \text{MAE} = \frac{\sum_{i = 1}^n|a - a_{pred}|}{n}
\end{equation}
over all thrust coefficients and all simulations $n = \text{14}$. While $a$ is the induction evaluated from a simulation at a specific acceleration at a given $C_T$, $a_{pred}$ is the predicted induction for this acceleration either based on the thrust-induction curve obtained from the uniform case or from the thrust-induction curve corrected by the acceleration term. 
Since the power coefficient is calculated as ${C_P = C_T(1-a)}$, the MAE directly translates to a variation in $C_P$ for a given $C_T$. Results are presented in Fig. \ref{fig:fig_4} (b). If no correction is considered, the mean error increases with an increasing thrust coefficient up to a value of around 0.021. Correcting the predictions for flow acceleration yields a reduction of error for all thrust coefficients. The maximum error in this case at high thrust is approximately 0.004 which corresponds to a mean error reduction of around 80~\%. At this point, it is important to note two things. First, the MAE was only calculated for the presented simulations. It does not yield insights into specific cases, but rather gives an estimate for the potential error reduction. Second, the length scale $l$ was set to one without further investigation. Fine-tuning of it has the potential to further reduce the error, as will be investigated later.

\section{Discussion}
\begin{figure}[t]
    \centering
    \includegraphics[width=8.5cm]{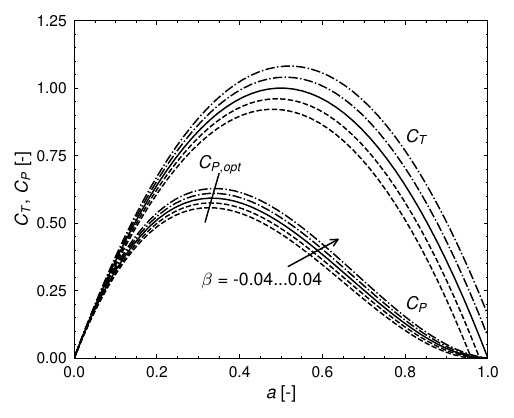}
    \caption{$C_T$ and $C_P = C_T (1-a)$ based on Eq. \eqref{eq:fe} in dependency of the induction for various $\beta$ and $l = \text{1}$.}
    \label{fig:fig_5}
\end{figure}

\noindent The model results indicate that the general approach of a linear superposition of the flow acceleration on the induction in the wake is valid up to a certain degree. It seems as if in an accelerating flow field, the acceleration can be interpreted as an additional force that is proportional to the induction. 

\subsection{Influence on power performance}
\noindent Since power can be generally calculated as the product of thrust and velocity, the modification of the relationship between these two also yields a different power performance. The question is, whether this also alters the optimal operational point. This can be answered by solving for the optimal induction based on Eq. \eqref{eq:fe} and $C_P = C_T(1-a)$. For a constant $l$ this yields 
\begin{align}
  a_{opt} = \frac{2}{3} + \frac{1}{3}l\beta  - \frac{1}{3}\sqrt{1 + l \beta + l^2\beta^2}.
\end{align}
In case of a vanishing acceleration, $a_{opt} = \frac{1}{3}$, which is the classical result from momentum theory. However, for a non-vanishing acceleration, the optimal induction changes in dependency of the acceleration and the length scale. This is visualized in Fig. \ref{fig:fig_5}, where power and thrust coefficient curves for various accelerations are visualized. Not only the optimal induction changes, but also the maximum power coefficient is altered. In a flow with a positive acceleration, the power coefficient increases relative to the non-accelerating case and in a flow with a negative acceleration, it decreases.

\subsection{On the length scale}\label{subsec:theoptimallengthscale}
\begin{figure}[t]
    \centering
    \includegraphics[width=8.5cm]{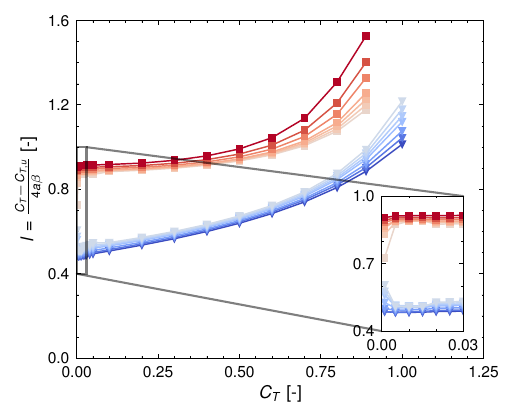}
    \caption{Extracted length scale $l$ from measured thrust-induction curves.}
    \label{fig:fig_6}
\end{figure}
\noindent Up to now, it was assumed that the length scale $l$ has a constant value, which was adopted from literature \citep{crespo_SurveyModellingMethods_1999,dar_AnalyticalModelWind_2022}. However, it is actually possible to determine it for the presented simulations by solving Eq. \eqref{eq:totct} for $l$. This is visualized in Fig. \ref{fig:fig_6} where $l$ is plotted over the thrust coefficient. It can be observed, that $l$ depends on both the thrust coefficient and the velocity gradient. The latter seems to be particularly dominant at high thrust. It is remarkable, that two separate branches of $l$ develop, depending on whether the acceleration is negative or positive. The fact that the branch resembling the negative acceleration cases takes higher values than the branch with the positive acceleration cases at a given thrust coefficient reflects the previously observed phenomenon that the impact of flow acceleration is higher in case of a negative acceleration than in case of positive acceleration. 
Very close to zero thrust, the curves converge more or less to a value of around 0.7, although the general behavior in this region seems to be highly non-linear. 
The previously employed value of 1 D is in the range of values $l$ takes within the full range of thrust coefficients. In the light of the prediction results presented earlier, this value can be seen as a good compromise between low and high thrust cases.
One could start now to interpret a lot into these results. However, one needs to keep in mind that $l$ is a variable in a simple, linear model and \mbox{Fig. \ref{fig:fig_6}} identifies it based on high-fidelity simulations, which per se do not necessarily act as linear as the model.
Thus, from \mbox{Fig. \ref{fig:fig_6}} the main conclusion from an engineering perspective is that one could probably find an even better fit than shown in Fig. \ref{fig:fig_1} by differentiating between positive and negative acceleration and possibly also including a thrust-dependency. 

\subsection{Limitations and future work}

\noindent As very idealized simulations were performed, a natural limitation is that the model still needs to prove its general applicability for more complex flow cases. As such, the atmospheric flow over complex terrain or the flow within a dense wind farm come to mind. In these scenarios, it is likely that the streamwise velocity gradient is not constant. 
To apply the presented model in these cases, one could interpret Eq. \eqref{eq:u1inf} as a linearization of the velocity field around the hub height velocity. Then, $l$ could yield an indication where the velocity behind the AD position needs to be evaluated in the undisturbed flow field. However, by now it is not clear if this approach would work in practice. 

As seen in Fig. \ref{fig:fig_5}, the optimal point of operation changes when a turbine is subjected to an accelerating flow field. This gives rise to the question, if existing control strategies can cope with this or if it would result in the turbine operating in a suboptimal state with respect to both, loads and power performance; a question which should be addressed in future works. 

Lastly, only scenarios of a comparably low induction \mbox{($a<$ 0.5)} were investigated numerically. Thus, it is of interest how the model performs at high-induction scenarios. Within this context, one could consider extending recent work on momentum theory \citep{liew_UnifiedMomentumModel_2024}.

\section{Conclusions} 
\noindent Streamwise acceleration of the background flow field is usually not considered for wind turbine performance predictions. In this work, it was shown that neglecting this effect can have a significant impact on induction and thus aerodynamic performance predictions. A simple model, which extends classical momentum theory to the case of a constantly accelerating streamwise flow field, was presented. Without any parameter tuning, the model yields an error reduction by eighty percent, and results indicate that fine-tuning of the model could yield an even higher prediction accuracy. The model shows that the induction at which maximum power performance is reached and the maximum power performance itself are influenced by flow acceleration, implying that wind turbines might not operate optimally under such conditions. 

\section*{Code and data availability}
\noindent EllipSys3D used for the simulations is a proprietary software developed at DTU Wind and Energy Systems and distributed under licence. The data used in this paper are publicly available at the following DOI: \url{https://doi.org/10.11583/DTU.27222912} 

\section*{Author contributions}
\noindent CPZ derived the model, performed the simulations and drafted the article. CPZ, NT and MG contributed to idea, methodology, and analysis and reviewed and edited the manuscript.

\section*{Competing interests}
\noindent
The contact author has declared that none of the authors has any competing interests.

\section*{Acknowledgements}
\noindent
We gratefully acknowledge the computational and data resources provided on the Sophia HPC Cluster at the Technical University of Denmark DOI: \url{https://doi.org/10.57940/FAFC-6M81}

\printbibliography

\end{document}